# Response to the IUPAC/IUPAP Joint Working Party Second Report
*'On the Discovery of Elements 110-118'*

A. Marinov[1], S. Gelberg[1], D. Kolb[2] and G. W. A. Newton[3]

[1] Racah Institute of Physics, The Hebrew University of Jerusalem, Jerusalem 91904, Israel
[2] Department of Physics, University GH Kassel, 34109 Kassel, Germany
[3] Heron's Reach, 382 Mossy Lea Road, Wrightington, Lancashire, WN6 9RZ, UK

*Abstract:* Back in 1971 fission fragments were observed by us in Hg sources separated from two CERN W targets that were irradiated with 24 GeV protons. The masses of the fissioning species were measured and heavy masses like 272, 308 and 317-318 were found and interpreted as due to the superheavy element 112 with 160-161 neutrons and various molecules of it. Based on the measured mass of the produced superheavy nucleus cold fusion reactions like $^{88}Sr + {}^{184}W \rightarrow {}^{272}112$ and $^{86}Sr + {}^{186}W \rightarrow {}^{272}112$ were deduced. The ordinary heavy ion reaction $^{88}Sr + {}^{184}W$ has been studied and characteristic X-rays of element 112 and a very high-energy $\alpha$ particle in coincidence with a fission fragment have been observed. The data have been studied by the TWG and they were concerned about the question of the reaction mechanism since very large fusion cross sections, in the region of a few mb, have been deduced in the secondary reaction experiments. This question can now be answered in view of our recent discovery of long-lived super- and hyperdeformed isomeric states. The JWP did not accept our arguments and this response answers their queries.

The second report of the JWP regarding our Z = 112 work is based on their first report. For the sake of clarity we consider both reports in our response.

Our claim for discovering element 112 has already been considered by the TWG [1]. It was based, in one set of experiments, on the observation of fission activities *with measured appropriate masses*[1] in species which followed the chemistry of Hg, and like the latter electroplated without applying voltage on Cu and evaporated at about 300° C[2]. Furthermore, from the measured masses the cold fusion reaction $^{88}Sr + {}^{184}W$ was deduced, and in another experiment this reaction has been studied and evidence for characteristic X-rays of element 112[3] and for a very high energy (12.16 MeV) $\alpha$-particle[4] in coincidence with a fission fragment were found. (According to the kinematics only the isotopes $^{271}112$ and $^{272}112$ could have been produced in this cold fusion reaction). Altogether we have found about 100 of these heavy Eka-Hg atoms, and the usual background in the measurements was zero. To the best of our knowledge such a quality of data has never been obtained in the discovery of other superheavy elements, where the identification was usually based on just a few atoms. The data were described in Refs. [2-8].

---

[1] Recently, in the 3rd Int. Conf. on Exotic Nuclei and Atomic Masses, the recent data of the GSI and Dubna groups were presented. In the summary talk [21] given by Dr. C. N. Davids of ANL he mentioned that what he thinks is missing in their data are mass measurements. Such essential measurements were performed by us about 30 years ago and consistently interpreted back in 1984 [4].

[2] This low temperature eliminated any element with $90 \leq Z \leq 111$.

[3] Photons with respective energies that are within 330 eV and 160 eV equal to the predictions for $K_{\alpha 1}$ and $L_{\beta 4}$ X-rays of element 112 have been seen in coincidence with low-energy particles [5 – 8], while the corresponding differences for adjacent elements are 4 keV and 800 eV respectively. The particles were assumed to be protons. Long-lived proton radioactivity has been observed by us. (See down below and Ref. [13]).

[4] The highest known today.

As mentioned above these data have been considered by the TWG in 1992 [1]. Generally speaking their attitude toward our work was positive. The TWG "…judges the experiments so interesting that it will express hope that they will get a follow-up" (quote from the minutes of the TWG meeting in Mogilany-Krakow 1-5 July 1991), or: "In your beautiful work which has been carried on during so many years, you have found, I think, many unexpected results…" (quote from a letter of Professor M. Lefort, a member of the TWG, to A. Marinov of March 20, 1991). They did not question the validity of our experimental data, but they were concerned about the problem of the reaction mechanism. Even then they pointed out that in both experiments "…this possibility cannot be definitely dismissed", or "…cannot be said to be impossible. Further work is needed". At that time only a partial answer could have been given by us to this question. First, we have claimed that from the experimental point of view it was shown that actinides, and in particular $^{236}$Am and $^{236}$Bk (in isomeric states), have been produced by secondary reactions in the same W target [9]. If $3.1 \times 10^5$ atoms of $^{236}$Am and $4.4 \times 10^4$ atoms of $^{236}$Bk have been produced in the target then the production of about 500 atoms of $^{\sim272}$112 is possible. This is particularly so since hot fusion reactions with large competition with fission are responsible for the production of the actinides while very cold fusion with much less fission is responsible for the production of element 112. Secondly, it was shown [4] that in the secondary reactions the projectiles are fragments that were produced just within about $5 \times 10^{-14}$ s before interacting with another W nucleus in the target. During this short time they are still at high excitation energy and quite deformed. Deformations have a very strong effect on the fusion cross sections between heavy nuclei as demonstrated by the well-known sub-barrier fusion effect [10,11] and seen in Fig. 4 of Ref. [7]. In addition, it was pointed out to the TWG that, like in the actinides [9], perhaps a long-lived isomeric state, rather than the ground state, was produced in element 112. However, at that time we did not have any clue about the character of the isomeric state, and whether it might help to explain the large cross section.

It should be mentioned that these arguments have been considered favorably by the TWG. In the above mentioned letter sent to A. Marinov by Professor M. Lefort on March 20, 1991, he said: "The possibility of producing long lived isomeric states in neutron deficient very heavy nuclei is indeed reasonable, as well as the hypothesis of production of highly excited deformed fragments as possible projectiles".

Continuing our research, at the time when the first JWP was established we were able to propose a more complete answer to the cross section problem. Based on our discovery [12,13] of the existence of long-lived high-spin superdeformed isomeric states,[5] it was suggested that similar isomeric states were formed in element 112 (as well as in the actinides). It is clear that much less penetration and dissipation is needed in order to produce the compound nucleus in the superdeformed shape rather than in its normal states. Hence much less or even no extra-push energy is needed and the fusion cross section is expected to be much larger.

---

[5] The evidence for the existence of these isomeric states is based on the observation of a relatively low energy and very enhanced alpha particle group where the enhancement is in accord with penetrability calculations for a superdeformed to superdeformed transition, and the alpha particles themselves are in coincidence with superdeformed band gamma ray transitions, and on the observation of long-lived proton radioactivities.

In their first report [14] the JWP did not follow the TWG but rather questioned the validity of the data themselves. They say (under **112;** 04 [14]): "The situation pertaining to these collaboration results has not changed substantially since the TWG judgment. If anything, it has become weaker because independent attempts to duplicate the process of fusion with secondary residues from high-energy proton irradiations of heavy targets have failed to find yields of elements more than a half dozen atomic numbers greater than that of the target (71Ka01, 73Ba01, 73Ge01) rather than the three dozen or more invoked by Marinov et al."

This statement is not correct: As mentioned above, in a study that was performed over about 15 years [9], $^{236}$Am and $^{236}$Bk (in long-lived isomeric states) were identified in the W target. These are 21 and 23 atomic numbers greater than that of the target. Besides, the TWG report was issued in 1992, and it is logically impossible to claim that since then the situation became weaker because of the above quoted papers that were published about twenty years earlier, and were known to the TWG. As a matter of fact, the quoted papers were already been dealt with by us back in 1984 [4] and it was shown that they could not prove the point made by the JWP.

In the second assessment of the JWP first report (under **112;** 08) [14] addressing our discovery of the long-lived superdeformed isomeric states which can explain how the production of element 112 by us could be possible, they say: "These two papers continue to press arguments for the existence of very long-lived isomeric states of actinides and transactinides and of very high fusion cross-sections for their formation, each several orders of magnitude beyond current understanding. These extraordinary phenomena are, in part, necessary for the acceptance of the collaborations' interpretation. The JWP remained unmoved."

This harsh verdict goes against the judgment of the TWG as expressed in the last above quotation from Prof. Lefort letter to A. Marinov of March 20, 1991.

It seems that there are clear inconsistencies between the assessments of the TWG and of the JWP.

Since our first submission to the JWP of our claim for priority in discovering element 112, further work was carried out by us regarding the existence of the long-lived isomeric states. The results have been summarized in two comprehensive papers [15,16][6] and have been submitted to the JWP for their second report [19]. In [16], strong evidence for the existence of a long-lived high-spin hyperdeformed isomeric state is given. It is based on the observation of a 13 orders of magnitude retarded (40 d $\leq$ $t_{1/2}$ $\leq$ 2.1 y) high-energy α-particle group of about 8.6 MeV in coincidence with superdeformed band γ-ray transitions, where the energy of the α-particles fits with theoretical predictions for a $III^{min} \rightarrow II^{min}$ transition. (In addition several more long-lived superdeformed isomeric states were observed. A summary of all the new transitions is given in Table 3 of Ref. [16]). In [15] a quantitative interpretation is given to both the low energies and the enhanced lifetimes of the unidentified α-particles seen in various actinide sources separated from the CERN W target in terms of $II^{min} \rightarrow II^{min}$ and $III^{min} \rightarrow III^{min}$ transitions.

Thus, long-lived isomeric states exist not only in the second minimum of the potential but also in the third minimum, and the evidence for this is based on 16 α - γ coincidence events where the γ-rays fit beautifully as superdeformed band transitions (Fig. 11 in Ref. [16]), and on several hundreds of low energy α-particles (Figs. 4-5, in Ref. [15]) where the background in these measurements is about zero.

---

[6] Short summaries of these papers are given in Refs. [17,18].

Like in the first report, the attitude of the JWP towards our work remains negative in its second report as well [19]. Their assessment is essentially the same as described above in their first report.

As has already been mentioned above our claim for the existence of very long-lived isomeric states of actinides and transactinides, of very high fusion cross sections for their formation, is not "beyond traditional understanding". Long-lived isomeric states have been produced by the $^{16}$O + $^{197}$Au [12,13] and $^{28}$Si + $^{181}$Ta [16] reactions, and also by secondary reactions, in $^{236}$Am and $^{236}$Bk [9] and in heavier actinide nuclei [15]. Superdeformed long-lived high spin isomeric states have also been predicted by Nilsson et al. back in 1969 [20], and similar effects of this type can exist also in the hyperdeformed region.[7] There is no reason to assume that long-lived high spin states could not be produced in the superheavy element region as well. The long measured half-life of the observed fission activity of several weeks and the large observed fusion cross sections indicate that this is indeed the case. Regarding the argument of the JWP about the fusion cross sections, we have already mentioned that the combined effect of having deformed fragments ("projectiles") and producing the compound nucleus in the super- or hyperdeformed isomeric state increases the fusion cross section by many orders of magnitude.

The JWP further writes: "As indirect evidence, their discovery of long-lived $^{236}$Bk and $^{236}$Am more than a decade ago is frequently cited in their papers, yet the several existing compendia of isotopes do not acknowledge the existence of these species."

The evidence for the existence of the long-lived isomeric states in $^{236}$Bk and $^{236}$Am is based on measuring during about 15 years the decay curve and half-life of a 5.76 MeV α-particle group from radioactive sources which followed the specific chemistry of Bk and Am, respectively [9]. Identification of an isotope on the basis of its α-decay energy and half-life is a standard procedure. In the "Table of Isotopes" by R. B. Firestone et al. there is a reference to our work of Ref. [9]. For both $^{236}$Am and $^{236}$Bk our work "87Ma21" is quoted under "Populating Reactions and Decay Modes". For $^{236}$Bk it is also mentioned that the type of the reaction is W(p,x). Our results regarding the isomeric states in $^{236}$Am and $^{236}$Bk and their formation via secondary reaction in a W target irradiated with 24-GeV protons are also quoted in the "Evaluated Nuclear Structure Data File" (ENSDF) by F. Orlando.

The JWP further writes: "The collaboration results include mention of observing long-lived proton-decay, of deformed spallation products undergoing secondary fusion reactions, and of hyperdeformed shapes any of which significant topics by themselves should have attracted studies by other groups years ago. Yet this has not occurred."

All our measurements repeated themselves several times and their statistical significance has been checked. We are not responsible for other groups' research programs.

The JWP further writes: "The collaboration's arguable use of forceful expressions such as "overwhelming evidence", "clear and proven", and "impossible to refute" is neither convincing nor swaying. Extraordinary intriguing phenomena, not much selective in their measured character, are, in part, necessary for the acceptance of the collaborations' interpretations of their data. The Joint Working Party needs much more to be able to relinquish its deeply felt unease that the tautological rationalization of the Marinov *et al.* measurements remains inadequate."

---

[7] As shown in [15], due to calculations of Howard and Möller [24], hyperdeformed states actually could be the true ground states in the heavy actinide nuclei and in the superheavy element region.

We do not think that we exaggerate when we describe our data with the above quoted expressions. Observation of fission fragments from sources that followed the chemistry of Hg [2,3], measuring the masses [4] of the fissioning nuclei and characteristic X rays of element 112 [5-8], are indeed, to our mind, very convincing. We also think the same about our discovery of the long-lived super- and hyperdeformed isomeric states [12,13,15,16] and that it is not justified to ignore them. In addition, as was mentioned above, a coincidence event between a 12.16 MeV α particle and a fission fragment has been seen in the study of the heavy ion $^{88}$Sr + $^{184}$W reaction [5 – 8] where, from the kinematics point of view, only $^{271}$112 and $^{272}$112 isotopes could have been produced in the reaction. This measurement is of the correlated type as the events studied by the GSI group, except that in our case the background was zero (Fig. 3 in [6] and [7] and Figs. 7 and 8 in [8]), and the coincidence time was 1 μs as compared to correlation times of milliseconds to tens of seconds. Why the GSI experiment is considered by the JWP as producing "high-quality data with plausible interpretation" and ours is completely ignored?

Finally in the SUMMARY OF JWP01 CONCLUSIONS [19] it is written: "Also, despite efforts by the Marinov *et al.* collaboration using atypical studies in conjunction with speculative theory to re-enforce their claim to element 112, we maintain that the results of secondary interactions involving hyperdeformed long-lived products of long lifetime and high production probability remain unconvincing curiosities, all aspects of which warrant more selective investigation."

Nothing in our theoretical explanations is speculative: standard penetration calculations including deformation up to super- and hyperdeformation have been used by us.
Already the TWG expressed that: "The possibility of producing long lived isomeric states in neutron deficient very heavy nuclei is indeed reasonable, as well as the hypothesis of production of highly excited deformed fragments as possible projectiles" (the above mentioned letter of Professor M. Lefort from 1991). The several weeks-long measured lifetime of the fission activity in the Hg sources shows that a long-lived isomeric state was produced in element 112. Long-lived super- and hyperdeformed isomeric states have been discovered by us using the $^{16}$O + $^{197}$Au [12,13] and $^{28}$Si + $^{181}$Ta [16] reactions, and it was shown (Fig. 8 in Ref. 15) that the production of a very heavy compound nucleus in a super- or hyperdeformed isomeric state is much more probable then its production in a normal state. Also the increased cross section as a result of deformations of the projectile and/or target is a well-known effect. The measured long lifetime of Z=112 and the large fusion cross section strongly indicate that super- or hyperdeformation is involved.

It also seems to us that the JWP should have said what is more selective than fission fragments from separated Hg sources followed by mass measurements of the fissioning nuclei, characteristic X-rays of element 112 and a coincidence measurement between an α particle and a fission fragment, determination of isotopes according to their chemical behavior and α-particle energies and lifetimes. Or, what is more selective than enhanced α-particles where the enhancement fits with penetrability calculations for superdeformed to superdeformed transition and the α-particles themselves are in coincidence with superdeformed band γ-ray transitions. Or, what is more selective than abnormally high energy and very retarded α-particles in coincidence with superdeformed γ rays, where the high energy is in accord with predictions for hyperdeformed to superdeformed transition, or, in addition, low energy and very enhanced α-particle groups, where both the energies and the enhancements fit with hyperdeformed to hyperdeformed transitions.

Before concluding let us comment on a question that was raised recently, namely, is it justified to expect that element 112 will act like Hg, since some relativistic calculations indicate that it might show properties more like a noble gas [22].

First let us mention that already in 1971 we took into account the possibility that element 112 may be more volatile than Hg. (Similar to the fact that Hg is more volatile than Cd.)[8] Secondly, fact is that the fission fragments were seen preferentially in the Hg sources and not in the Au, Tl and Pb sources [23], and that they also were seen in the mass separator experiments, where a Hg source first was electroplated on Cu without applying any voltage and then was evaporated in the ion source at a low temperature which eliminated any element with $90 \leq Z \leq 111$. It is clear that the measured fission activity basically followed the chemistry of Hg, otherwise one would have readily lost it in the complex chemical procedure. Only Z=112 including relativistic effects as has been calculated by Pershina et al. [22] is similar enough to do this.[9] Therefore, our chemistry did isolate element 112 and no other. It is essential that in our case the chemical separation was done on Hg and element 112 at various oxidation states and not at an elemental state like in [25,26] where one is basically sensitive to volatility and adsorption properties of an element. Furthermore, as mentioned in footnote 9 and is important for our experimental procedure, according to the relativistic calculations [22] the binding energies of Z=112 and Hg on Cu are almost the same.

In summary let us mention a point of principle. One has to distinguish between the experimental evidence for the existence of an element and the understanding of the way of its production. Already in ancient times people knew and admired when they held a piece of gold in their hand. They of course did not have any idea about supernova explosions and how the gold or the heavy elements in general were produced. In the modern history of physics one may recall that when the continuous spectrum of β rays was discovered, people were so confused that they were even willing to give up the conservation law of energy. Yet, no one had doubt about the very existence of the continuous β spectrum itself.

In this respect our discovery of element 112 was done back in 1971 [2,3] by observing the fission fragments in Hg sources and by measuring the fissioning masses. The understanding of the masses of the fissioning nuclei in 1984 [4] strongly support the original results. The measured masses led to a reasonable hypothesis about the fusion reactions that took place in the W targets, and by a study of a similar but ordinary heavy ion reaction evidence for characteristic X-rays of element

---

[8] For instance, in the measurement of the energies of the fission fragments [2,4] the source was cooled to liquid nitrogen temperature to avoid its evaporation.

[9] It is shown that the $p_{1/2}$ and $s_{1/2}$ level energy distance even increases from Hg to element 112 and thus make partial $p_{1/2}$ occupancy as improbable as in Hg. Occupation of the $p_{1/2}$ shell by at least one electron (Z=113 and higher) or at least one electron less in the s-d shell (Z below 112) would drastically change the chemical and physical properties. E.g. from the experience with s-d atomic level structures in the periodic table we see that the evaporation temperatures stay high as long as the s-d shell is not closed, with a strong drop for the closed $d^{10}s^2$ configuration as in Hg and Z=112. On the other hand the $ns_{1/2}$ and $(n-1)d_{5/2}$ which were quite apart in Hg are now almost energetically degenerate, because the $7s_{1/2}$ orbital becomes more bound due to a direct relativistic effect and the $6d_{5/2}$ less bound due to indirect relativistic effects (stronger shielding of the nuclear charge by the relativistically enhanced deeper binding of the $s_{1/2}$ and $p_{1/2}$ orbitals). The less bound $6d_{5/2}$ makes it more reactive and the more bound $7s_{1/2}$ makes it more noble than in Hg. These two effects may compensate and make chemical behaviors of Hg and element 112 similar. (E.g. according to [22] the binding energies ($D_c$) of Hg and element 112 on Cu are almost the same). However the chemistry of Hg cannot be similar to that of any other superheavy element in this Z region.

112 and for a very high energy α particle in coincidence with a fission fragment was obtained [5-8]. Our discovery of the long-lived super- and hyperdeformed isomeric states [12,13,15,16] enables us also to understand, in a fully consistent manner, the production of element 112 in both the secondary and the ordinary heavy ion reactions. Based on all this evidence we believe that element 112 has been discovered by us back in 1971.